# Helical and Straight Solitons Induced by Vortex Beams with Off-Axis Pivots in Cubic-Quintic Nonlinear Media

Jing Chen[1], Rongcao Yang[1,2,*], and Boris A. Malomed[3,4]

[1] *College of Physics and Electronics Engineering, Shanxi University, Taiyuan 030006, China*

[2] *Shanxi Key Laboratory of Wireless Communication and Detection, Taiyuan, 030006, China*

[3] *Department of Physical Electronics, School of Electrical Engineering, Faculty of Engineering, Tel Aviv University, Tel Aviv 69978, Israel*

[4] *Instituto de Alta Investigación, Universidad de Tarapacá, Casilla 7D, Arica, Chile*

* *sxdxyrc@sxu.edu.cn*

**Abstract**: We address the existence and dynamics of solitons propagating along helical (spiral-rotating) and straight trajectories in the bulk waveguide with the cubic-quintic nonlinearity and a single- or multi-ring potential. The input is taken as a vortex beam with a single or multiple phase singularities (pivots), displaced from the waveguide's axis. The single off-axis pivot creates a one-ring helical soliton, with a closed or open (split) ring structure. The vortex beams with multiple off-center pivots give rise to complex states with multi-ring, multi-core, and/or multi-split structures. The multi-core/split solitons propagate along helical channels, or straight ones, which are parallel to the propagation axis. The sign and period of the helicity can be precisely adjusted by applying a torque with an appropriate angular velocity.

## 1. Introduction

Vortex beams are waves carrying orbital angular momentum (OAM), which feature a helical wave front circulating around the phase singularity (vortex pivot), as demonstrated in diverse settings [1-14]. Vortex beams offer a significant application potential in areas such as optical measurements, light-field modulation, high-resolution microscopy, quantum data processing, as well as micro/nano-manipulations and fabrication by means of optical technologies [15–18].

The azimuthal instability of vortex solitons in Kerr (cubic) nonlinear media usually initiates the onset of splitting and collapse [1, 9], while the azimuthal stability can be provided by the competing cubic-quintic (CQ) [19] or quadratic-cubic [20] nonlinearities. A fundamental prediction is that vortex-soliton solutions, alias vortex rings, with any integer topological charge (winding number) $m$, produced by the two-dimensional (2D) nonlinear Schrödinger equation (NLSE) with the CQ nonlinearity, are stable against spontaneous splitting in the free space (in the absence of an external potential), provided that their total power (see Eq. (3) below) exceeds a certain critical value [19, 21]. It is found that in nonlinear metamaterial waveguides, the purely cubic nonlinearity leads to the splitting of vortex beams into annular soliton clusters (filamentation), while competing cubic-quintic nonlinearities suppress this instability and stabilize the vortex dynamics [22]. This prediction was partly confirmed in the experiment [23]. The confinement and guidance of vortex beams [24] in the CQ medium and the creation, decay, and regeneration of vortex clusters [25] were also observed experimentally. In addition, elliptical and rectangular solitons [26], super-Gaussian beams exhibiting self-trapping [27], bright/dark vortex solitons with high topological charges [28, 29] and cosh-Gaussian beams evolving into flat-top solitons [30] were investigated theoretically. Stable vortex solitons were predicted as well in the framework of the fractional NLSE with the CQ nonlinearity [31].

The use of external potentials is an alternative method for the stabilization of vortex solitons. Recent theoretical studies have demonstrated stable propagation of multipole vortex solitons, including vortex dipoles, quadrupoles, sextupoles, and octopoles, in rotational *PT*-symmetric potentials [32] and in a rotating harmonic trap [33]. Stable higher-charge vortex quantum droplets in a ring potential [34] and multi-ring nested vortex solitons in an axisymmetric radially-periodic lattice potential were predicted too [35].

Recently, interest has been drawn to vortex beams with a high topological charge, which normally exhibit a completely axisymmetric annular intensity distribution [29, 36-38]. On the other hand, experiments demonstrate vortex beams with an off-axis placement of their pivot – in particular, due to alignment errors [39]. Such an off-axis phase singularity breaks the symmetry of the beam, leading to new features of vortex solitons, which have been the subject of many studies [40-43]. For example, the generation of multi-off-axis circular Pearcey-like vortex beams in the free space was experimentally demonstrated [40], and off-axis chirped circular Pearcey-Gaussian vortex beams were theoretically investigated in a fractional system [41]. Such off-axis structured beams share similarities with sinh-Gaussian and cosh-Gaussian pulses produced by a decentered Gaussian superposition in nonlinear media [44, 45]. Further, off-axis vortex Gaussian beams were theoretically explored in strongly nonlocal nonlinear media, showing an overall orbital motion [42]. Multiple off-axis vortex beams were addressed in nonlinear Kerr media, exhibiting various controllable shapes with high topological charges [43].

However, the generation mechanism and distinctive dynamics of off-axis vortex beams in CQ media with concentric ring potentials remained unexplored.

In this paper, we investigate the existence, stability, and dynamics of helically and straightly propagating solitons in the framework of the CQ NLSEs with concentric ring potentials, and present unique characteristics that have not been observed in previous works, including the formation of stable open ring-core structures, the coexistence of helical and straight propagation channels for multi-core and multi-split states, and the tunability of the helicity sign and period via external angular velocity. Specifically, the OAM and motion of the phase singularity position of such vortex solitons are analyzed too. Vortex solitons with multiple phase singularities are also explored, featuring multi-ring and multi-core, closed or split, configurations, which propagate along helical (rotating spiral) channels or straight ones.

## 2. The model

The propagation dynamics of optical beams in the CQ medium with an external potential is governed by the respective NLSE [46, 47]:

$$i\frac{\partial \Psi}{\partial z} = -\frac{1}{2}\left(\frac{\partial^2 \Psi}{\partial x^2} + \frac{\partial^2 \Psi}{\partial y^2}\right) - |\Psi|^2 \Psi + |\Psi|^4 \Psi + V(x,y)\Psi, \tag{1}$$

where $\Psi(x,y,z)$ is the complex amplitude of the electromagnetic field. $z = Z/L_d$ is the propagation distance, normalized by the diffraction length $L_d = k_0 r_0^2$. $x = X/r_0$ and $y = Y/r_0$ are the transverse coordinates scaled to the characteristic scale $r_0$. A colloidal suspension of silver nanoparticles in acetone provides a realistic platform for this model with competing cubic and quintic nonlinearities [24]. Physical parameters at which the present setup can be implemented in the experiment is estimated as: the pump intensity 3 GW/cm$^2$ at the carrier wavelength 532nm, the beam waist 18 μm, and the sufficient propagation distance 10 mm.

The concentric ring potential

$$V(x,y) = -V_0 \sum_{j=1}^{k} \exp\left[-\left(r - r_j\right)^2 / d_j^{\,2}\right] \tag{2}$$

is constructed as a set of $k$ nested ring-shaped troughs with depth $V_0>0$, $r=\sqrt{x^2+y^2}$ is the radial coordinate, $r_j$ and $d_j$ standing for the radius and width of each ring, respectively. Such potential can be realized experimentally by computing the corresponding phase distribution and encoding it onto a spatial light modulator by dint of the computer-generated holography [24]. Equation (1) conserves the integral power (energy flow), OAM, and Hamiltonian [48], respectively:

$$\begin{aligned} P&=\iint\left|\Psi(x,y)\right|^2 dxdy, L=\iint \Psi^*\left(x\partial y-y\partial x\right)\Psi dxdy, \\ H&=\iint\left[\frac{1}{2}\left(\left|\Psi_x\right|^2+\left|\Psi_y\right|^2\right)+V(x,y)\left|\Psi\right|^2-\frac{1}{2}\left|\Psi\right|^4+\frac{1}{3}\left|\Psi\right|^6\right]dxdy. \end{aligned} \tag{3}$$

For stationary vortex states with phase $m\theta$, where $\theta$ is the angular coordinate and $m$ is the integer winding number (vorticity), the normalized value of OAM-per-photon is $L/P=m$ [2]. However, $L/P$ may be non-integer for spiral fractional vortex beams [49] and rotating vortex solitons [50].

To proceed with the analysis, stationary solutions of Eq. (1) are sought for as

$$\Psi(x,y,z)=\psi(x,y)\exp(i\beta z), \tag{4}$$

where $\psi(x,y)=\psi_r(x,y)+i\psi_i(x,y)$ is the transverse complex wave function, and $\beta$ is a real propagation constant. Inserting this in Eq. (1) leads to the stationary equation:

$$\beta\psi=\frac{1}{2}\left(\frac{\partial^2\psi}{\partial x^2}+\frac{\partial^2\psi}{\partial y^2}\right)+\left|\psi\right|^2\psi-\left|\psi\right|^4\psi-V(x,y)\psi, \tag{5}$$

which can be numerically solved using the squared-operator method or Newton-conjugate-gradient methods [51], to produce vortex solitons with the above-mentioned phase $m\theta$.

Generally, phase singularities of vortex beams are centered on the optical axis, however, alignment errors or the intention to construct new configurations may displace the singularities from the central position [39] and thus break the axial symmetry of the beam's intensity distribution [33, 52], significantly affecting the propagation of the vortex beams [40-43].

To address this possibility, we introduce an initial guess for the solution of Eq. (5) for the vortex multi-soliton multi-ring solution with amplitude $A$ and off-axis displacements as follows:

$$\psi(x,y)=A\sum_{n=1}^{l}\exp\left[-\frac{(r-q_n)^2}{p_n^{\ 2}}\right]\exp(im_n\theta_n), \tag{6}$$

where $q_n$ and $p_n$ are the radial position and width of the *n*-th ring, $m_n$ is the topological charge of the respective vortex beam, $\theta_n=\arctan\left[(y-\rho_n\sin\phi_n)/(x-\rho_n\cos\phi_n)\right]$ being the angular coordinate defined with respect to the displaced center of the *n*-th vortex, in terms of the global polar coordinates ($\rho_n$, $\phi_n$), and $l$ is the number of off-axis vortices in the initial ansatz, independent of the potential ring number $k$ in Eq. (2).

The stability analysis of the vortex beams can be performed by adding perturbations to the stationary solution as follows: $\Psi(x,y,z)=\left[\psi(x,y)+v(x,y)\exp(\lambda z)+\bar{w}(x,y)\exp(\bar{\lambda}z)\right]\exp(i\beta z)$, where $v$ and $w$ are infinitesimal perturbations and $\lambda$ is the complex instability growth rate. The substitution of the perturbed expression in Eq. (1) and linearization leads to the eigenvalue problem [51],

$$\frac{1}{2}\left(\partial_x^2+\partial_y^2\right)v-V(x,y)v-\beta v+2|\psi|^2\left(v-\psi^2w\right)+|\psi|^2\left(w-3\bar{\psi}^2v\right)=-i\lambda v \tag{7a}$$

$$-\frac{1}{2}\left(\partial_x^2+\partial_y^2\right)w+V(x,y)w+\beta w+2|\psi|^2\left(\bar{\psi}^2v-w\right)+\bar{\psi}^2\left(3\psi^2w-v\right)=-i\lambda w \tag{7b}$$

where the overbar stands for the complex conjugate. This linear system can be solved by means of the Fourier collocation method [51]. The vortex soliton is stable if all eigenvalues are pure imaginary.

## 3. Results and discussions

The configurations of the ring potential in Eq. (2) are shown in Figs. 1(a)-1(c), including the single-ring, double-ring, and triple-ring potentials. Taking the single-ring case as an example, we derive the linear eigenvalue spectra versus the potential depth $V_0$ from the numerical solution of the linearized version of Eq. (5), as shown in Fig. 1(d). Here we set $x$, $y \in [-40, +40]$ and the number of grid points $N$=512 to produce the linear eigenvalue spectra. It is clear that, as the potential depth increases, the discrete eigenvalues shift upward. The linear eigenvalue spectrum for the single-ring potential with $V_0 = 1$ is presented in Fig. 1(e), where the red-solid-dot eigenvalues with marks 2 and 3 correspond to the two degenerate linear eigenstates in Figs. 1(f) and 1(g). The two linear eigenstates $\psi_{1,1}$ and $\psi_{1,2}$ can be superposed to construct linear vortex states with topological charges $m=\pm1$ as $\psi_{m=\pm1}=\psi_{1,1}\pm i\psi_{1,2}$ .

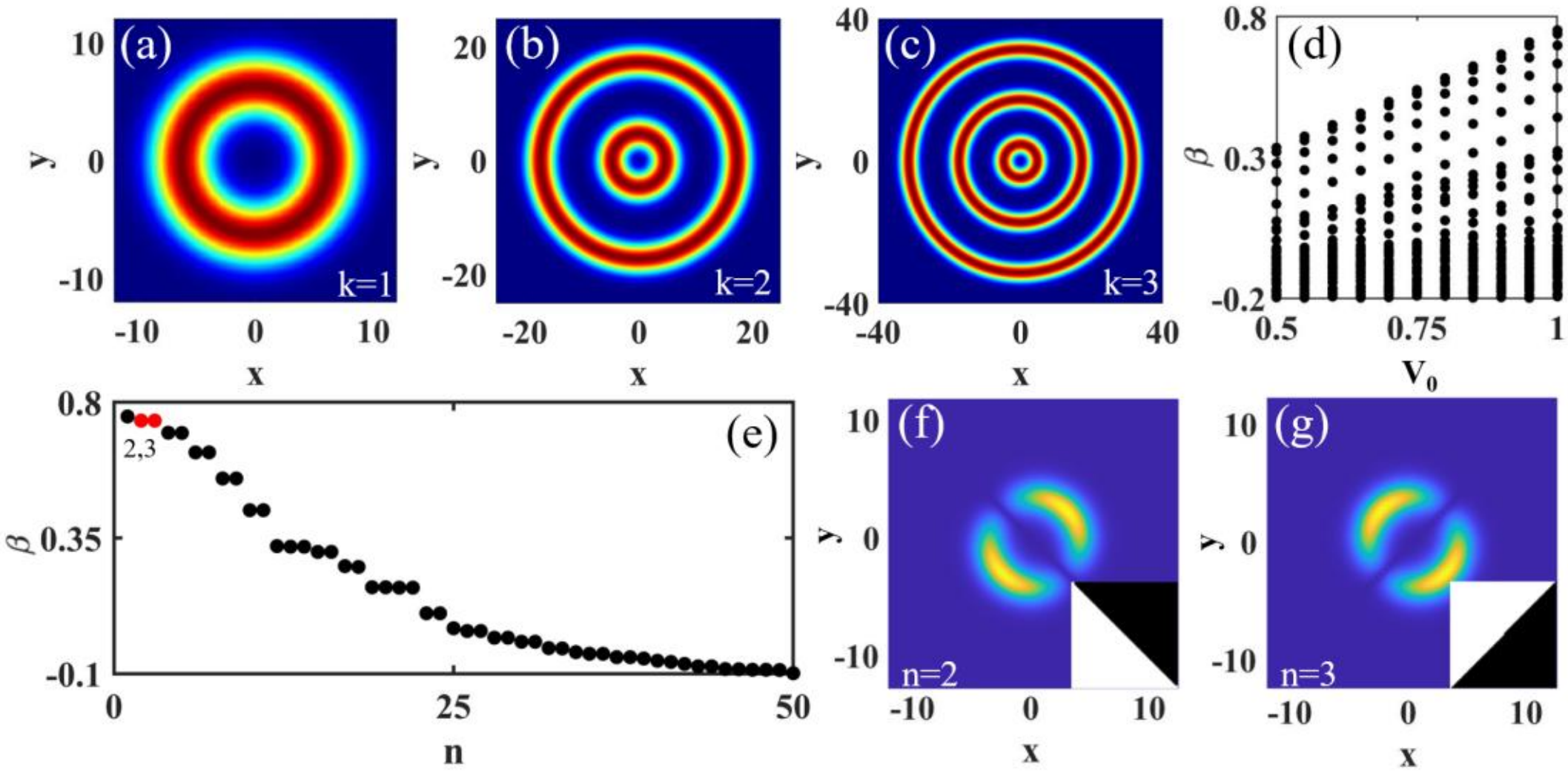

Fig. 1. The concentric ring potential with $V_0$=1 for (a) $k$ =1, $r_1$=2π, $d_1$=2.5; (b) k=2, $r_1$=1.5π, $r_2$=5.5π, $d_{1,2}$=2.5; (c) $k$=3, $r_1$=1.5π, $r_2$=5.5π, $r_3$=10π, $d_{1,2,3}$=2.5. (d) Linear eigenvalue spectra of the single-ring potential with varying potential depths $V_0$. (e) The typical linear eigenvalue spectrum at $V_0$ = 1. (f, g) Degenerate linear eigenstates corresponding to the red-solid-dot eigenvalues in (e), with their phase in the bottom-right panel.

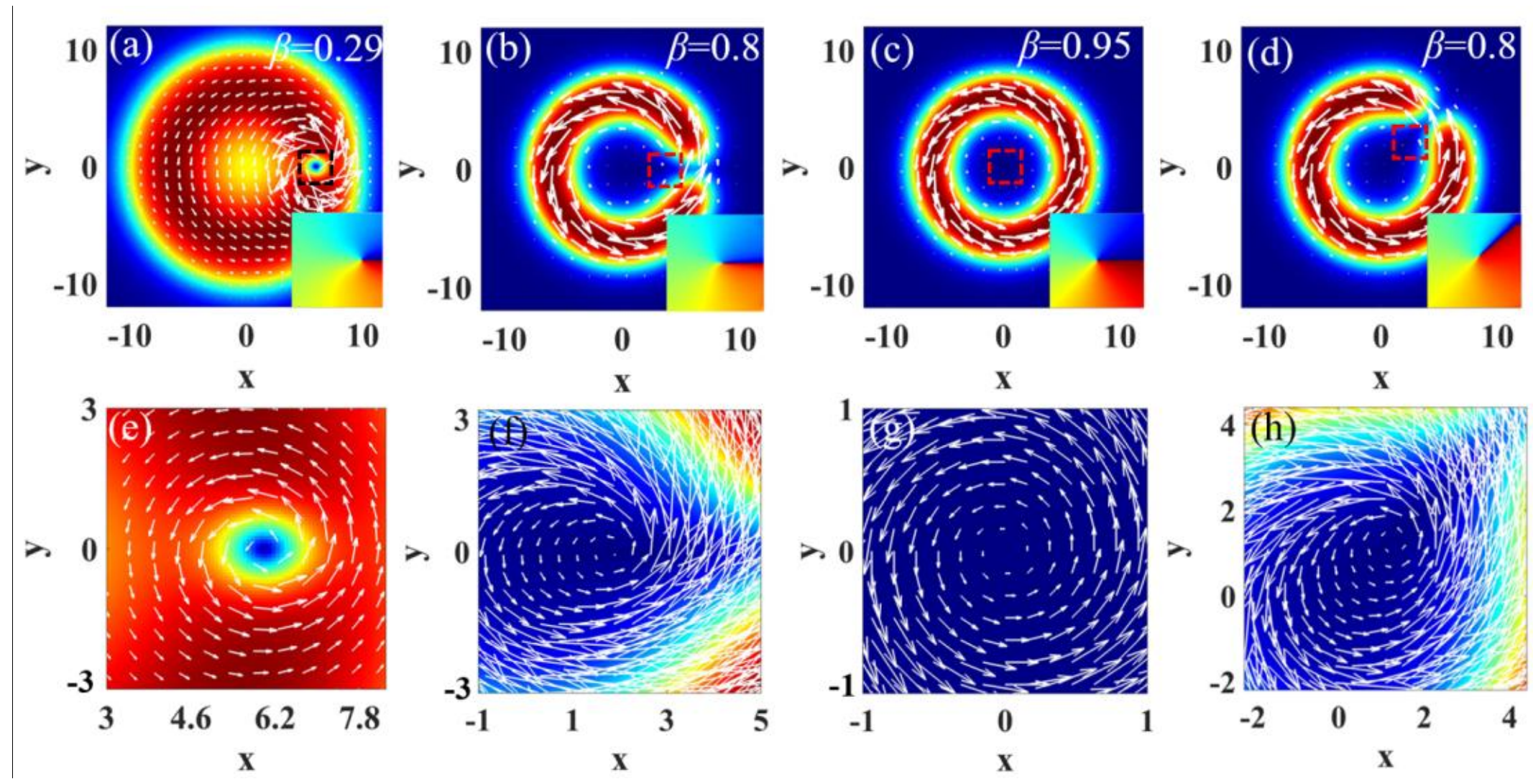

Fig. 2. The field's amplitude, Poynting vector, and phase structure of the single-ring vortex soliton with the topological charge $m$ = +1 and propagation constants (a) $\beta$ = 0.29, (b, d) $\beta$ = 0.8, (c) $\beta$ = 0.95. (e)-(h) The zoom of the dotted box in (a)-(d). The stability of these solitons is corroborated below by Fig. 4.

We start with the single-ring vortex solitons carrying topological charge of $m$ = +1, centered at an off-axis phase singularity. As shown in Fig. 2, for vortex solitons supported by the single-ring potential with $V_0 = 1$, the increase of the propagation constant leads to a gradual displacement of the position of the phase singularity towards the domain's edge, forming a dark core near the edge [Fig. 2(a)] or a split annulus with an opening [Fig. 2(b)] in the distributions of the field's amplitude. These outcomes are a consequence of the fact that the vortex' pivot (phase singularity), originally displaced from the center, breaks the rotational symmetry, resulting in a non-equilibrium intensity gradient force that propels the phase singularity into motion. For comparison, in the case without the initial off-axis displacement of the phase singularity, a conventional vortex soliton is produced with the central phase singularity [Fig. 2(c)]. Moreover, the dark core or opening of the single-ring vortex solitons can appear at any position on the ring [see, e.g., Fig. 2(d)]. We also plot the global [in Figs. 2(a)-2(d)] and local [in Figs. 2(e)-2(h)] energy-flow density, represented by the Poynting-vector field (white arrows), $\boldsymbol{S}=\left(\Psi\nabla\Psi^{*}-\Psi^{*}\nabla\Psi\right)/2$, with $\nabla$ standing for the gradient operator. Note that the energy flows in these vortex solitons keep the form of closed loops [Figs. 2(a)-2(d)], even in the split-ring states [Figs. 2(b) and 2(d)], which is a consequence of the local gradient-balance mechanism. The counterclockwise annular energy flow around the phase singularity can be clearly seen in the zoomed boxes [Figs. 2(e)-2(h)].

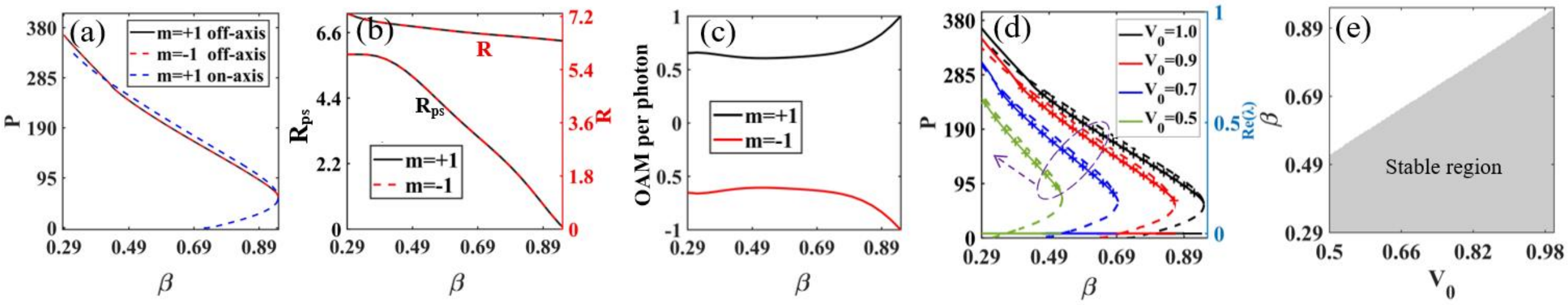

Fig. 3. (a) The power for on-axis (blue) and off-axis (black and red) single-ring vortex solitons with $V_0$ = 1.0. (b) The phase singularity (ps) position $R_{\text{ps}}$ and root-mean-square (rms) radius $R$, (c) OAM-per-photon vs. the propagation constant for the off-axis vortex solitons with $m$ = ±1. (d) The power $P$ and the instability growth rate Re(λ) for off-axis single-ring vortex solitons with $m$ = +1 and decreasing potential depth $V_0$ = 1.0, 0.9, 0.7, 0.5. Solid and cross-marked lines represent the off-axis closed- and split-ring structures, respectively; dashed line denotes the on-axis counterpart power. (e) The stability domain of the single-ring vortex soliton in the ($V_0$, $\beta$) plane is shaded gray, which coincides with its entire existence domain.

The family of the numerically produced off-axis single-ring vortex-soliton solutions is characterized by the dependences of their power, position of the phase singularity (pivot), root-mean-square (rms) radius $R=\left(P^{-1}\iint\psi^2(x,y)\left(x^2+y^2\right)dxdy\right)^{1/2}$ [53], OAM-per-photon on the propagation constant, and stability results, which are plotted in Figs. 3(a)-3(e). The position of the phase singularity $R_{ps}$ is defined as the distance from the coordinate origin to the pivot. As the propagation constant increases, the power of the off-axis single-ring soliton exhibits monotonous decay [Fig. 3(a)], along with the gradual contraction of the rms radius [Figs. 2(a)-2(c) and 3(b)] and the position of the phase singularity [Fig. 3(b)]. With the increase of the propagation constant, the OAM-per-photon of the vortex solitons with topological charge +1 gradually increases up to 1 at $\beta$ = 0.95. At this point, the phase singularity is located at the central position ($R_{ps}$ = 0) and the ring soliton restores its complete annular shape [Fig. 2(c)]. To investigate the characteristics of the off-axis vortex solitons, the power as a function of the propagation constant for the on-axis (central phase singularity) single-ring vortex solitons, carrying topological charge $m$ = +1, is presented in Fig. 2(a). This dependence is nonmonotonic owing to the competing cubic and quintic nonlinearities, with the upper and lower branches merging at the critical point $\beta$ = 0.95. The power curve of the off-axis family, which exists only on the upper branch above the critical point, eventually converges to the critical point. This indicates that the off-axis vortex family emerges through a bifurcation from the on-axis vortex family.

It is noted that the OAM-per-photon is non-integer at $\beta$ < 0.95, which originates from the asymmetric intensity distribution in the integration domain when the phase singularity is displaced from the beam's center. Thus, the deviation of the OAM-per-photon from the integer topological charge serves as a measure of the off-axis displacement, decreasing with the increase of the propagation constant [Fig. 3(c)]. The OAM-per-photon for the vortex solitons with topological charge -1 is simply the negative of that for +1.

The relation between the power and propagation constant of the on-axis and off-axis single-ring vortex solitons depends on the strength of the external potential, as illustrated in Fig. 3(d) for decreasing $V_0$ = 1.0, 0.9, 0.7, 0.5 at the fixed topological charge, $m$ = +1. It is seen that, for different potential depths, the power curves of the off-axis vortex-solitons family are located on the upper branches of the on-axis counterparts and converge to the corresponding critical power points. For a given propagation constant, a deeper potential corresponds to a higher power for the off-axis vortex solitons. Solid and cross-marked lines in Fig. 3(d) represent the closed- and split-ring structures of off-axis vortex solitons, respectively, and the dashed lines denote the on-axis ones. The Vakhitov-Kolokolov criterion is commonly applied to judge the stability of solitons [54]. However, it only provides a necessary condition. For vortex states, the criterion cannot capture the azimuthal instability that leads to spontaneous vortex splitting. Therefore, the eigenvalue spectrum is also calculated by means of the numerical solution of Eq. (7), as shown in Fig. 3(d), which demonstrates that the off-axis vortex solitons with different potential depths are stable within the entire regimes. The corresponding stability domain is shown as the shaded region in Fig. 3(e). Note that the vortex states are robust against moderate variations of the rings' parameters, which merely cause a slight shift of the propagation- constant interval, without altering the qualitative behavior. In the subsequent analysis, we set $V_0 = 1$.

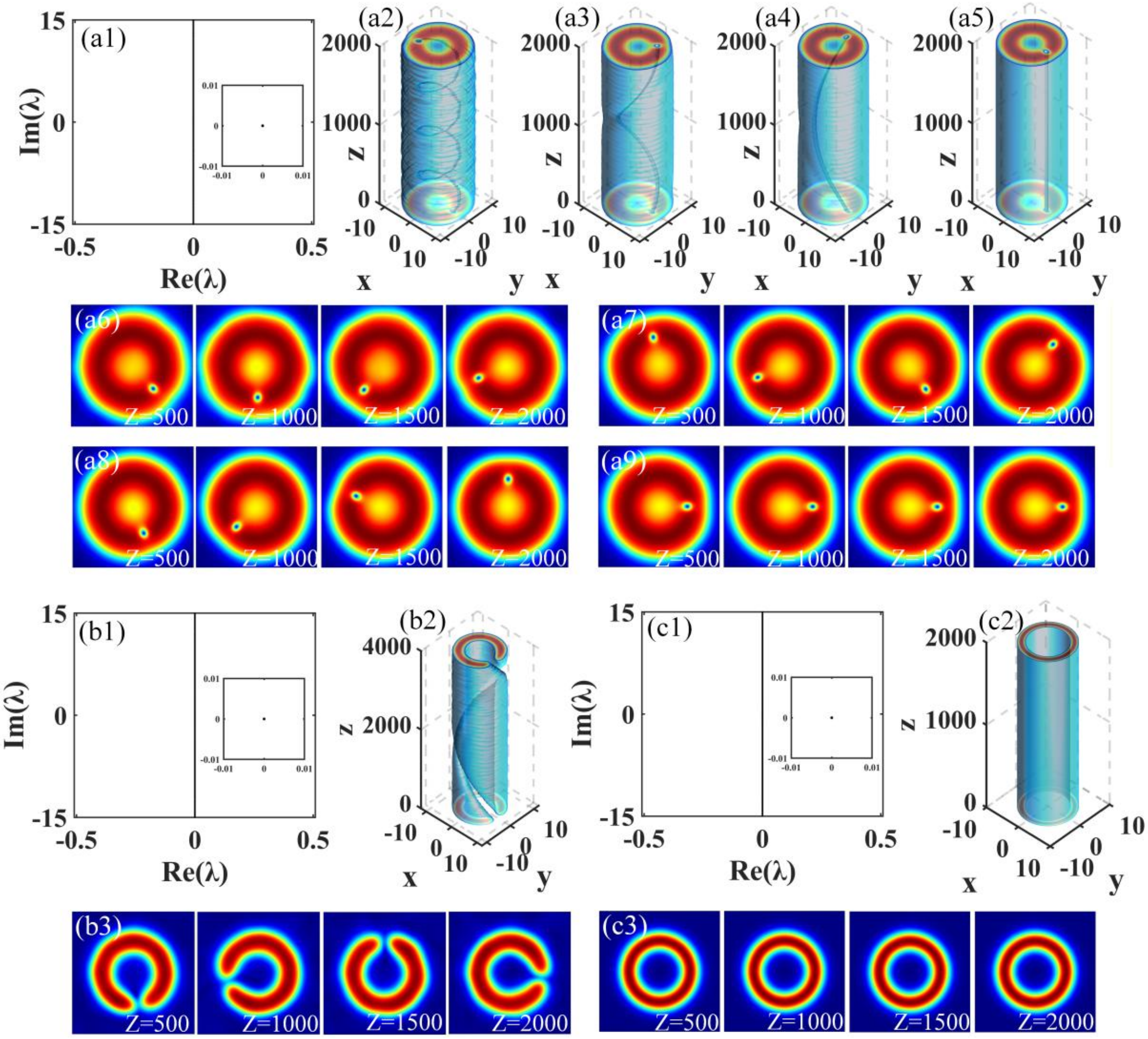

Fig. 4. Linear stability spectra and dynamics of off-axis vortex solitons: (a) $\beta = 0.29$, (b) $\beta = 0.8$, and (c) $\beta = 0.95$. Panels (a1), (b1) and (c1) display the linear-stability spectra with the zero eigenvalue singled out. (a2)–(a5): 3D isosurfaces for 3% (mode 1), 1% (mode 1), 1% (mode 2), and 0% perturbations, with corresponding cross-sections presented by sections (a6)–(a9). Panel (b2) shows the 3D isosurface for the 2% perturbation, with the corresponding cross-section shown in (3b). (c2) and (c3): the perturbed 3D isosurface and its cross-section.

Next, we present results of the linear-stability analysis for the off-axis single-ring vortex solitons. Figures 4(a)-4(c) display the dynamics of the stationary solitons introduced in Figs. 2(a)-2(c) under the action of different perturbations, along with the perturbation eigenvalue spectra. Pure imaginary eigenvalues indicate that the solitons are stable, which is corroborated by simulations of their propagation. It is seen that the stationary soliton exhibits different angular velocities of rotation under different perturbations, because the rotation is induced by the perturbation applied to the stationary solution. In the perturbation eigenvalue spectra, plotted in Figs. 4(a1)-4(c1), there is a double zero eigenvalue corresponding to two degenerate eigenmodes, which account for the soliton's rotation in opposite directions [Figs. 4(a3, a4)]. Furthermore, comparing Fig. 4(a2) (3% perturbation) to Fig. 4(a3) (1% perturbation) demonstrates that, naturally, a smaller perturbation produces slower rotation, and no rotation occurs for the unperturbed vortex soliton [Fig.4(a5)]. The rotational dynamics of the single-ring vortex soliton with $\beta = 0.29$ can be clearly seen from the cross sections [Figs. 4(a6-a9)] corresponding to Figs. 4(a2-a5). The eigenvalue spectra, propagation and cross-sections of the perturbed single-ring solitons with $\beta = 0.8$ and $\beta = 0.95$ are presented in Figs. 4(b) and 4(c), respectively. Clearly, the single-ring split (open-ring) vortex soliton [see Fig. 4(b2)] rotates along a clockwise spiral channel (in this example, driven by a particular zero mode). However, when the off-axis displacement of the phase singularity vanishes [Fig. 2(c)], the rotation channel is no longer observed [see Figs. 4(c2) and 4(c3)], as the split-ring vortex soliton transitions to a ring-shaped one at $\beta = 0.95$. Thus, we conclude that the rotation of the off-axis vortex soliton arises from the excitation of the zero modes by the applied perturbation.

Figures 5(a)-(c) present the vortex solitons with topological charge $m = +1$, supported by the double-ring potential. At $\beta = 0.4$, distinct dark cores appear in the inner and outer rings [Fig. 5(a)], both rings exhibiting split configurations at larger $\beta = 0.77$ [Fig. 5(b) and 5(c)].

The vortex solitons with topological charge $m = +1$, supported by the triple-ring potential, are displayed in Figs. 5(d) and 5(e), again revealing the independent location of the dark cores in different cores. Figure 5(f) displays the power of the double- and triple-ring vortex solitons as a function of the propagation constant. Similar to Fig. 3(a), the power of the multiple-ring vortex solitons gradually decreases with the increase of the propagation constant. Comparing Figs. 5(f) and 3(a), it is seen that the greater number of the potential rings (annular troughs) naturally corresponds to a greater power of the vortex solitons.

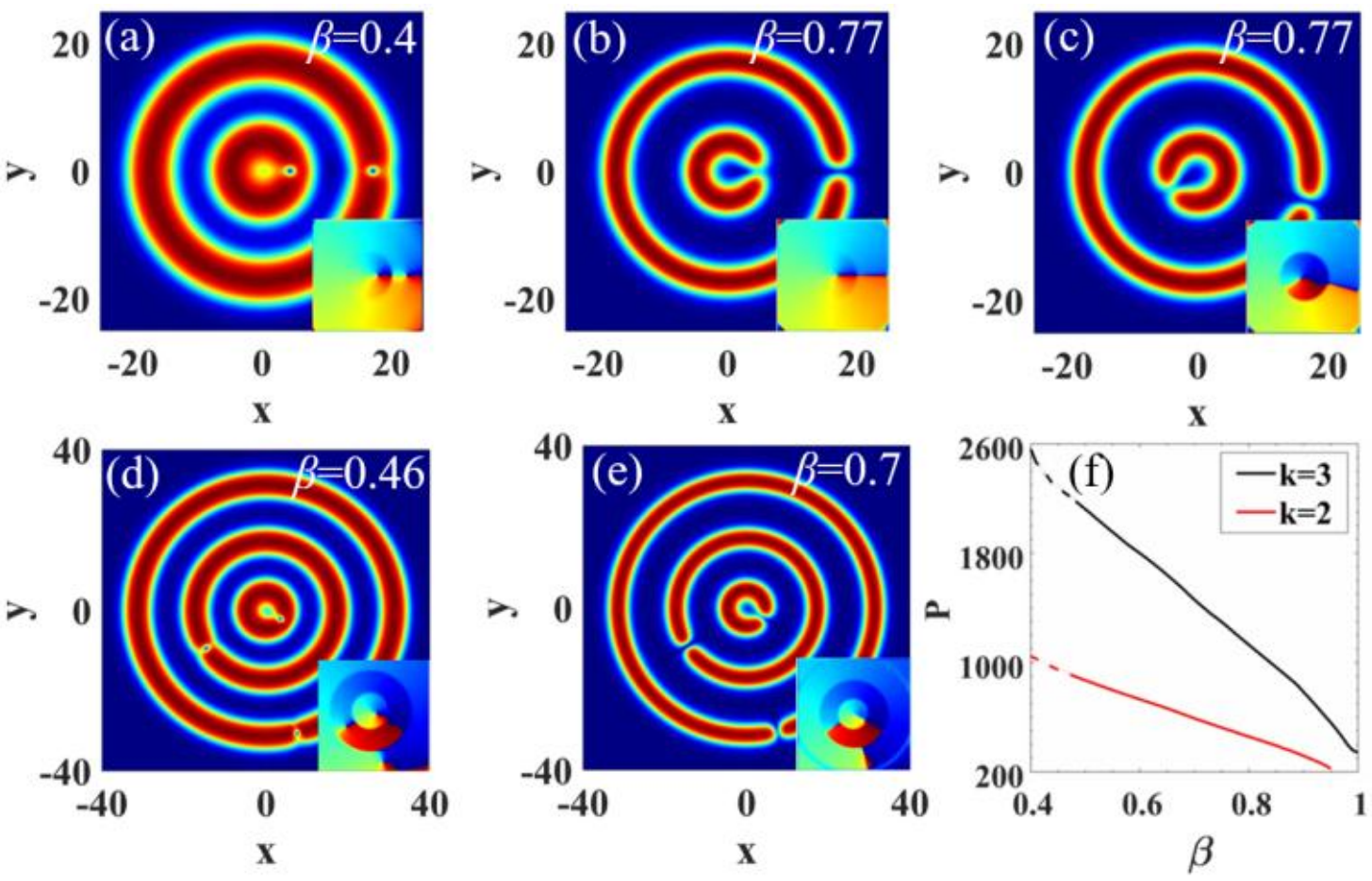

Fig. 5. (a)-(e) The field's amplitude and phase structure of the double-ring vortex solitons with (a) $\beta$ = 0.4 (unstable), (b, c) $\beta$ = 0.77 (stable), and triple-ring vortex solitons with (d) $\beta$ = 0.46 (unstable), (e) $\beta$ = 0.7 (stable). Panel (f) displays the dependence of the power on the propagation constant for the double- ($k$ = 2) and triple- ($k$ = 3) ring families, with the solid and dashed segments denoting stable and unstable solitons, respectively.

The double- and triple-ring vortex solitons may be unstable, as seen in Figs. 6(a) and 6(d). Typical examples of unstable and stable propagation, displayed in Figs. 6(b, e) and 6(c, f), corroborate the prediction of the linear-stability analysis. For the double- and triple-ring vortex solitons, the concentric rings with the dark-core structures demonstrate distorted transmission over long distances [Figs. 6(b) and 6(e)], which correspond to the positive instability growth rates in Figs. 6(a) and 6(d). However, the vortex solitons with the open-ring (split) structures remain stable in the course of the propagation [Figs. 6(c) and 6(f)], which have zero instability growth rates. It is observed that the rotation of the second ring is slower than that of the first ring (the rings are numbered from inside to outside as the first ring, second ring, etc.), as confirmed by the ($x$, $y$) cross-sections of the field-amplitude pattern, see Figs. 6(b2) and (c2). It is noted that reducing the inter-ring separation destabilizes the vortex solutions, eventually leading to structural collapse (not shown here in detail).

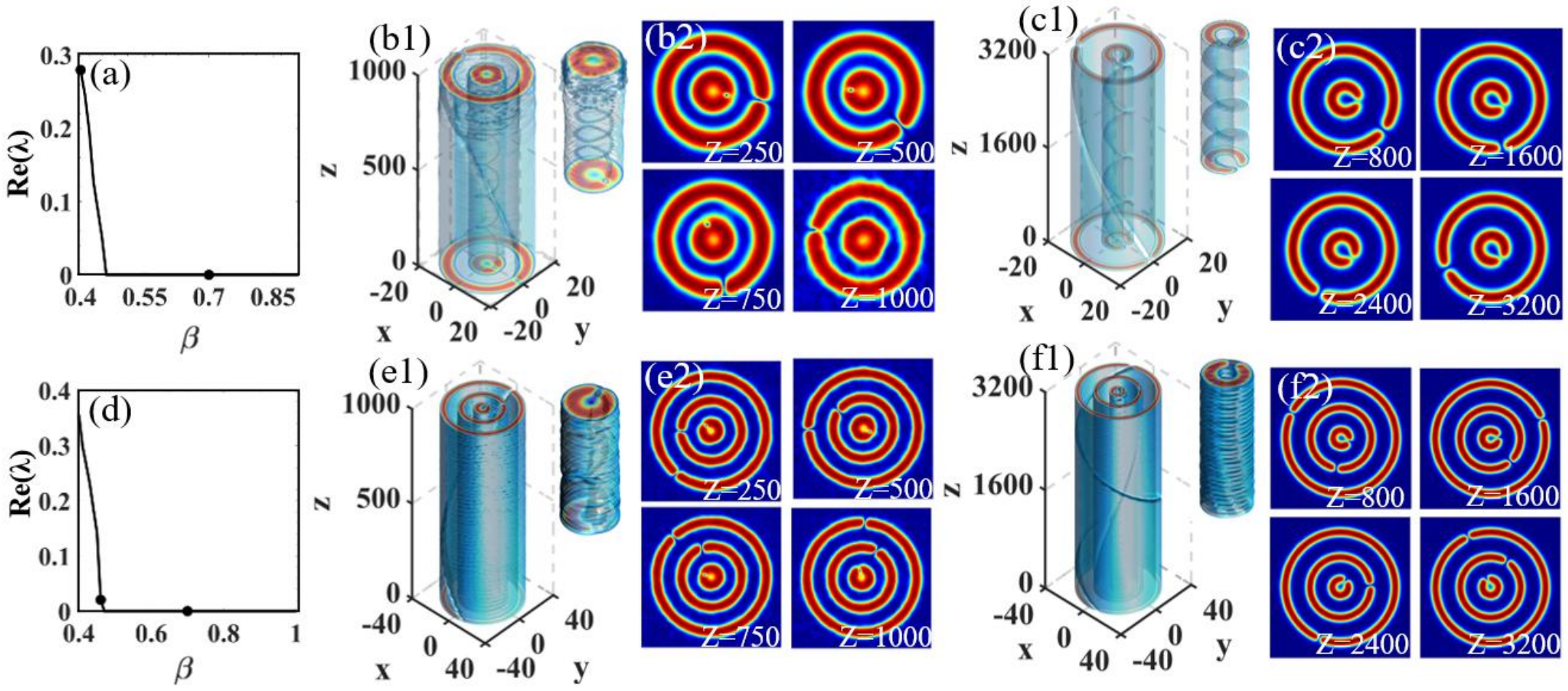

Fig. 6. (a) and (d) The instability growth rate vs. the propagation constant for the double- and triple-ring vortex solitons, respectively, which were introduced in Fig. 5. Panels (b, c) and (e, f) 3D isosurfaces and cross-sections of the field amplitude for the vortex solitons shown in Figs. 5(a, b) and 5(d, e), respectively. Insets in panels (b1, c1, e1, f1) display details in the first ring.

It is also interesting to consider vortex solitons hosting phase singularities with identical or opposite topological charges. Figures 7(a)-7(d) demonstrate the solitons with identical charges (+1), supported by the single-ring potential. Two phase singularities are located at diametrically opposite positions in the ring, forming a double-dark-core soliton for $\beta = 0.35$ [Fig. 7(a)]. Alternatively, in Fig. 7(b) the ring maintains the double split located on the $x$-axis for $\beta = 0.44$. In Fig. 7(c), both the power and OAM decrease with the increase of the propagation constant. Concurrently, a pair of two phase singularities keep the spatial symmetry while approaching each other [Fig. 7(d)].

In contrast, the vortex soliton bearing two opposite topological charges (±1) loses its vortex phase and forms a quasi-standing-wave pattern, building a dipole soliton with a split along the $y$ axis, as shown in Figs. 7(e) and 7(f). The phase distribution becomes a step structure alternating among $-\pi$, 0, and $\pi$. Naturally, OAM vanishes in this case, as shown in Fig. 7(g). It is also seen in Fig. 7(g) that the dipole-soliton's power decreases with the increase of the propagation constant. The two-vortex solitons are found to be stable in the entire range of values of the propagation constant.

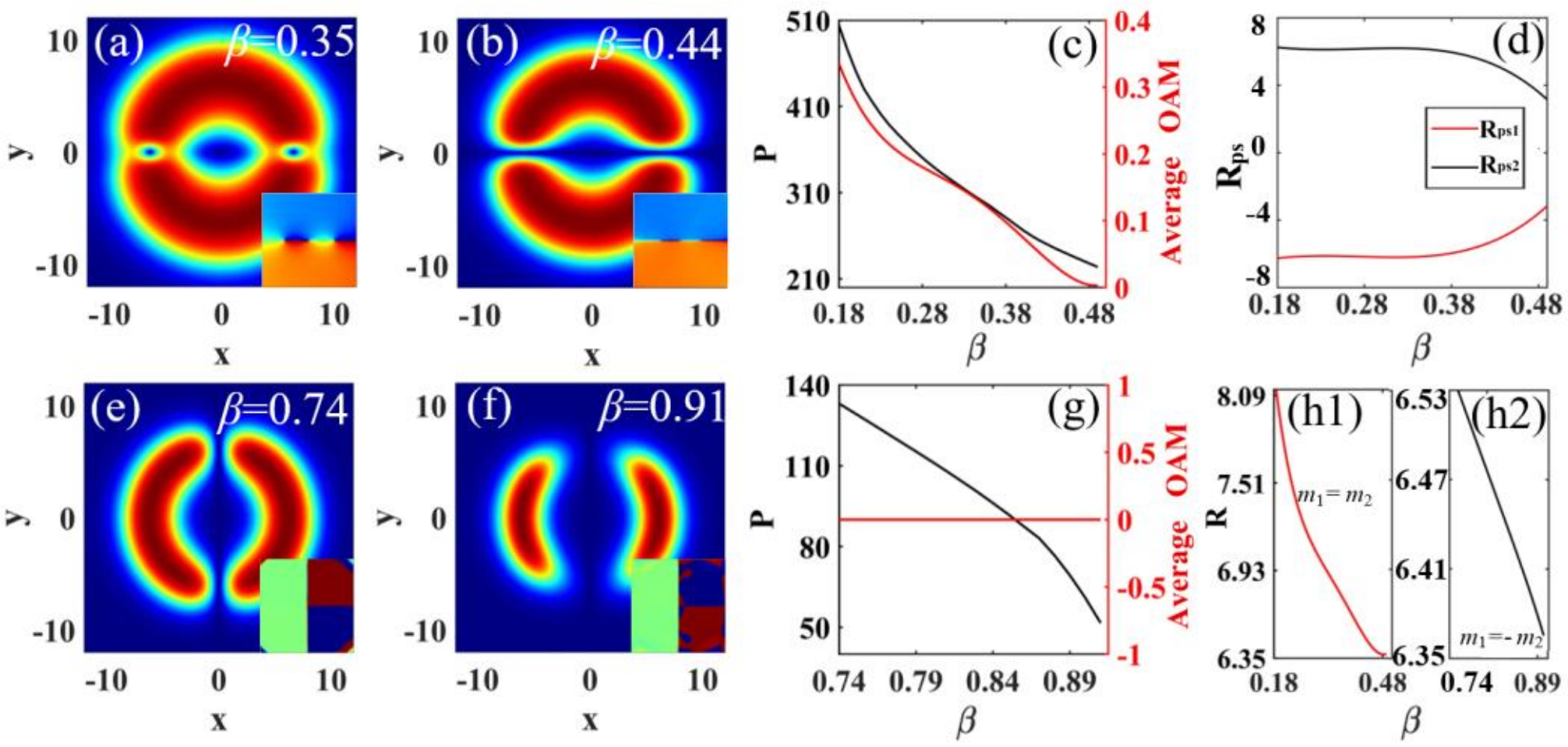

Fig. 7. The field's amplitude, phase structure, power, and OAM-per-photon, phase singularity position of the single-ring soliton with two (a)-(d) identical topological charges, $m_{1,2} = +1$, and (e)-(g) opposite topological charges, $m_{1,2} = \pm1$. (a) $\beta = 0.35$, (b) $\beta = 0.44$, (e) $\beta = 0.74$ and (f) $\beta = 0.91$. (h1, h2) The soliton's rms radii vs. the propagation constant.

Distributions of the soliton's field amplitude are similar for both cases of $m_{1,2} = +1$ and $m_{1,2} = \pm1$. In the intervals of $0.18 < \beta < 0.49$ and $0.74 < \beta < 0.91$, the soliton's radius decreases with the increase of the propagation constant, as seen in Figs. 7(h1) and 7(h2).

Figure 8 presents the instability growth rate versus the propagation constant, stability eigenvalue spectra, propagation simulations, and cross-sections for the single-ring solitons carrying identical or opposite pairs of the topological charges. A pair of mutually interwoven spiral channels, which represent the identical topological charges, arise in the course of the propagation of the single-ring vortex solitons, as illustrated in Figs. 8(c2, c3) and 8(e2, e3). Conversely, in the case of opposite topological charges, the phase singularities vanish, giving rise to the propagation of two parallel straight linear channels, as seen in Figs. 8(d2, d3) and 8(f2, f3). The stable helical and straight propagation channels may offer new degrees of freedom for light routing and topological state encoding [55, 56].

Figure 9 demonstrates the field's amplitudes and phase patterns of the vortex solitons with uniformly arranged multi-phase singularities, transforming from the multi-core vortex solitons [Figs. 9(a, b) and 9(e, f)] into the multi-split ones [Figs. 9(c, d) and 9(g, h)] with the increase of the propagation constant. This transition can be attributed to the decrease of the soliton power with $\beta$, which weakens the nonlinear refractive index and thereby reduces the self-trapping confinement on the vortex solitons, eventually allowing the multi-core structure to split structure. To construct these states, all topological charges are taken with the same sign, *viz*., $m_{1,2,3} = +1$, in Figs. 9(a)-9(d), and $m_{1,2,3,4} = +1$ in Figs. 9(e)-9(h). The field pattern is similar if all topological charges are taken to be -1. They repel each other in the CQ medium, which helps to stabilize the multi-vortex sets. With the increase in the number of vortex cores, more complex multi-core/split vortex solitons are generated, thereby creating multiple propagation channels. These findings are relevant to the field of optical data processing and may inspire new designs of transmission schemes for optical telecommunications.

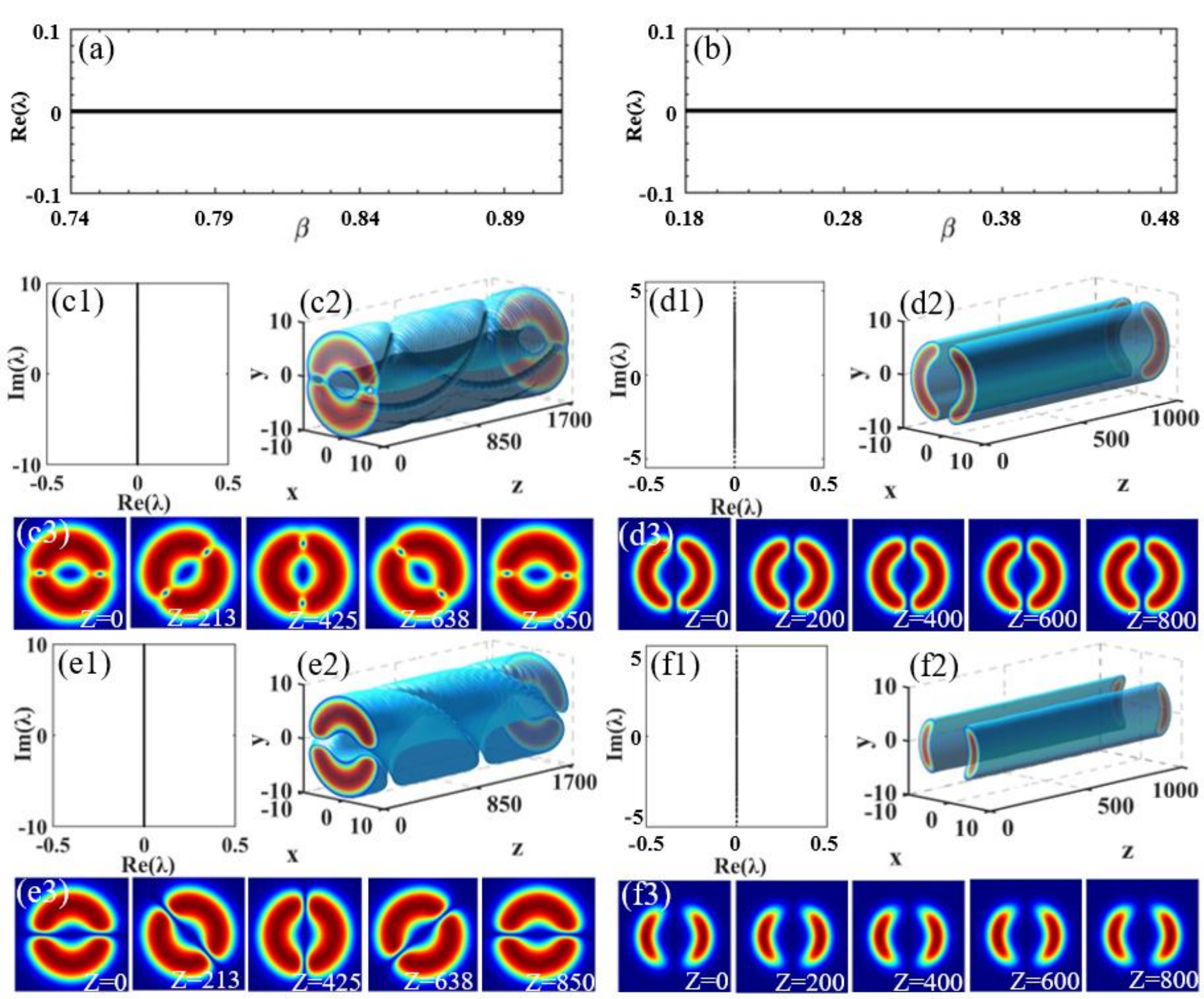

Fig.8. Single-ring solitons with two identical topological charges. (a) The instability growth rate vs. the propagation constant, (c, e) eigenvalue spectra, 3D isosurfaces, and cross-sections of the field-amplitude pattern at (c) $\beta = 0.35$ and (e) $\beta = 0.44$. (b, d, f) Same as (a, c, e), but for opposite topological charges at (d) $\beta = 0.74$ and (f) $\beta = 0.91$.

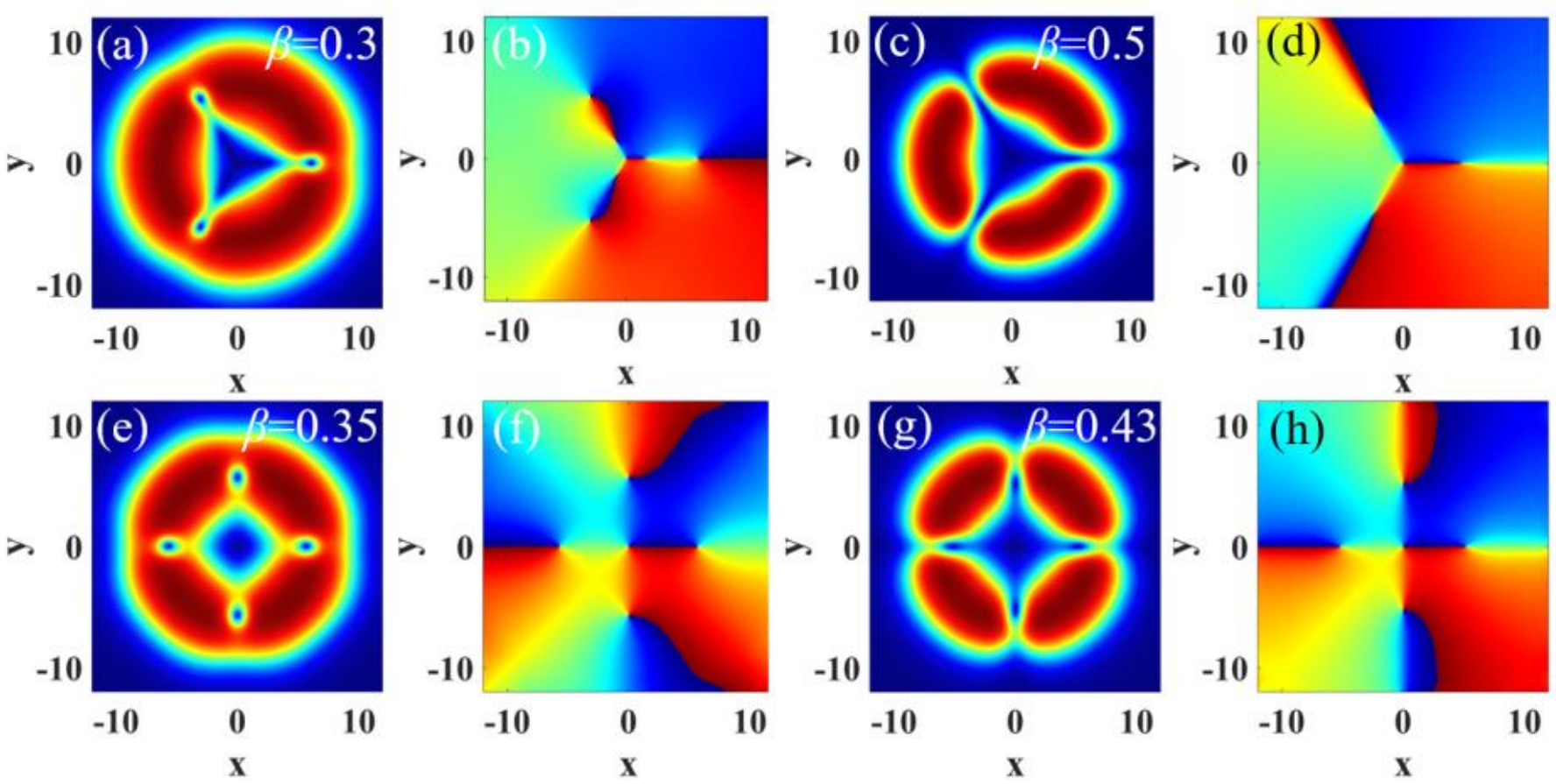

Fig. 9. The field's amplitude and phase structure of the single-ring vortex soliton with multiple phase singularities: (a, b) $\beta$=0.3, (c, d) $\beta$=0.5, (e, f) $\beta$=0.35, and (g, h) $\beta$=0.43.

Additionally, to precisely modulate the transmission period of the spiral channel, we imposed rotation with angular velocity Ω onto the solutions, taking $\Psi(x,y,z)=\psi(r,\theta-\Omega z)\exp(i\mu z)$ in the polar coordinates. Substituting this into Eq. (1) yields the Coriolis-force term $i\Omega(x\partial/\partial y-y\partial/\partial x)$, which drives vortex rotation with a period of $2\pi/|\Omega|$. For both the dark-core and split-single-ring vortex solitons, the rotation follows the same rule: a positive (negative) value of Ω causes counterclockwise (clockwise) rotation, with a period decreasing as |Ω| increases. Specifically, when small counterclockwise angular velocity Ω = 0.001 is applied to the dark-core vortex soliton, it maintains the counterclockwise propagation direction [Fig. 10(a)]. Increasing Ω to 0.002 enhances the Coriolis force and further reduces the rotation period [Fig. 10(b)]. Conversely, applying the same magnitude of Ω in the opposite direction (Ω = -0.001) reverses the propagation direction to clockwise [Fig. 10(c)], and, as the Coriolis force increases, the total angular velocity increases, while the period decreases accordingly, see Fig. 10(d). Similar helical propagation is observed for the split-single-ring solitons, as shown in Figs. 10(e)-10(h). The propagation period and direction of the vortex-soliton's helical channel can be altered through the appropriate selection of the parameters. Finally, we note that the inclusion of the rotational frequency Ω does not affect the above-mentioned field-amplitude patterns.

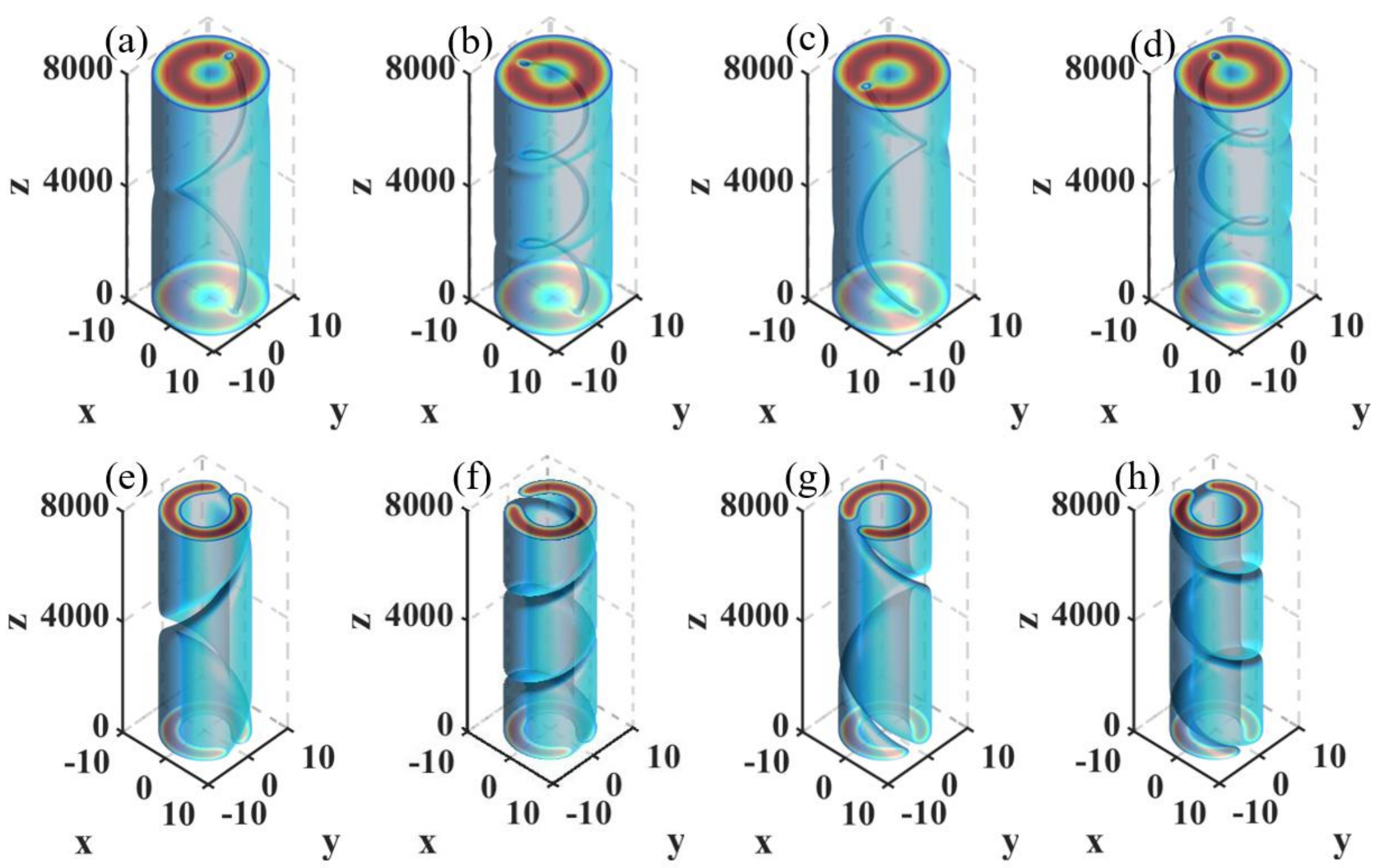

Fig. 10. (a)-(d) and (e)-(h) The dynamics of the dark-core and split single-ring vortex solitons shown in Figs. 2(a) and 2(b), generated by the application of rotation with angular velocity Ω. (a, e) Ω = 0.001, (b, f) Ω = 0.002, (c, g) Ω = -0.001, (d, h) Ω = -0.002.

## 4. Conclusion

We have investigated the propagation of multi-lobe vortex solitons along rotating and straight channels in the framework of the model governed by the nonlinear Schrödinger equation with the CQ (cubic-quintic) nonlinearity and single- or multi-ring (concentric) potential. We have demonstrated that the vortex soliton with the single off-axis pivot (phase singularity) can produce the vortex soliton in the form of the split ring, while multiple phase singularities in the concentric multiring array enable the formation of double- and triple-ring vortex solitons and double/three/four-(core/split) vortex solitons. In the course of the propagation, this multi-(core/split) configuration gives rise to two species of stable propagation channels, *viz*., the helical (rotating-spiral) and straight ones. The transmission direction and period can be precisely adjusted by the application of the additional angular velocity. The vortex solitons studied here may be useful for the design of optical data-transmitting and processing schemes, including light routing and topological state encoding.

**Acknowledgments**

This work was supported by the National Natural Science Foundation of China (Grant No. 62575165, 62305199), and Shanxi graduate Education Innovation Program (Grant No. 2024YZ06)

**Data availability**

Data may be made available upon a reasonable request.

**Declaration of competing interest**

The authors declare that they have no known competing financial interests or personal relationships that could have appeared to influence the work reported in this paper.

**CRediT authorship contribution statement**

**Jing Chen:** Writing – original draft, Methodology, Data curation. **Boris A. Malomed**: Writing – review & editing, Formal analysis, Conceptualization. **Rongcao Yang:** Supervision, Methodology, Funding acquisition, Conceptualization.